\documentclass[one column]{revtex4}
\usepackage{hyperref}
\usepackage{pdflscape}
 
\usepackage{subfigure}
\usepackage{epsfig}
 
\usepackage{amsmath,amsfonts}
\begin{document}

\title{{ Thermal fluctuations to thermodynamics of  non-rotating BTZ black hole }}
\author{Nadeem-ul-islam${}^{a}$}
  \email{drnadeemulislam@gmail.com} 
   \author{Prince A. Ganai${}^{a}$}
  \email{princeganai@nitsri.com}  
   \author{Sudhaker Upadhyay${}^{c,d}$}
 \email{sudhakerupadhyay@gmail.com; sudhaker@associates.iucaa.in}
 \affiliation{${}^{a}$Department of Physics,
National Institute of Technology,
Srinagar, Kashmir-190006, India }
   \affiliation{${}^{c}$Department of Physics, K.L.S. college, Magadh University, Nawada-805110,  India}
   \affiliation{${}^{d}$Inter-University Centre for Astronomy and Astrophysics (IUCAA) Pune, Maharashtra-411007}

\begin{abstract}
In this paper, we  discuss the effect of small statistical thermal fluctuations around the equilibrium
on the thermodynamics of small non-rotating BTZ black hole. This is done by 
evaluating the leading-order corrections  to the thermodynamical equation of states, namely,  entropy,  free energy, internal energy, 
pressure, enthalpy,  Gibbs free energy and specific heat  quantitatively.
In order to analyze the effect  of perturbations on the thermodynamics,  
we  plot various  graphs and compare corrected and non-corrected thermodynamic quantities with respect to event horizon radius 
of non-rotating BTZ black hole.
 We   also derive the first-order corrections to isothermal compressibility.
\end{abstract}
\maketitle

\section{Overview and motivations}
It  was  first  revealed  by  classical  theory  of  general  relativity (GR)  that  area of 
event  horizon  of  black  hole   never  decreases \cite{bh}.  There  was a clear  indication  that
thermodynamic  principles  can  be  incorporated  to  the  black  holes analysis.  In the thermodynamic  system, the  
entropy  plays a  central  role  to  study  the thermodynamic  properties  and  thermodynamic  evolution  takes 
place  in  such  a  way   that  entropy  never  decreases.  
 The  same  effect  was  observed   for the  area  of 
event  horizon   of  black  holes.  This  leads  to  a reasonable interpretation   for the  area  as an entropy  of  a  black  hole  and   we  can  see  that an  increase  of  area  of  event  horizon  is 
reminiscent  of  law  of   increase  of  entropy  of thermodynamic  system.  Also it was  found  that  surface gravity
$\kappa$   (which is measure of strength of gravitational field at the  event horizon) remains constant  over the event
horizon  surface.  This in turn  again  hints  the  reminiscence  with  zeroth  law  of thermodynamics,  which  states  
that  in  thermal  equilibrium, there  exists  a  common  temperature  parameter   for the whole  thermodynamic  system. 
It is quite  obvious  from  this  resemblance that  surface gravity  $\kappa$  of  black  hole  plays the role  of 
temperature in thermodynamic  system  \cite{la}. These analogies  compelled  Bekenstein  \cite {de}  to  propose  a
quantitative relation  between  entropy  and  area  of  event horizon of black hole.  But this  proposal  seems  to 
contradict  the second law of thermodynamics   as  any  thing  that  falls  in black  hole  could  not  escape  back. Consequently, it is
impossible  to  achieve  thermal  equilibrium  at  any  non-zero  temperature  between the  black  hole  and  any  
thermal radiation. However  this  puzzle  was  solved  by  taking    the  concept  of  quantum fluctuations into  account, 
which  amounts  to  the hawking  evaporation \cite{ab}.  Since black hole evaporates and
 it  loses the content (information) in it. 
As a result, the  Bekenstien entropy relation between the 
entropy and area 
also gets corrected with some leading-order  logarithmic    terms \cite{ cd}.

In order to study the leading-order corrections to thermodynamic  properties of black 
         holes,   several  attempts  have   been already made. For example, the  logarithmic 
         corrections to the entropy of  BTZ,  string-theoretic and rest other black holes, 
         whose  microscopic degrees of freedom are described by underlying conformal field 
         theory (CFT), have been found by
         using   Cardy formula  \cite{qw,er}. The correction terms to the entropy  are still 
         logarithmic in nature  while considering background matter field  into account 
         \cite{g1,s2,d3}. The logarithmic nature of 
 leading-order corrections to the entropy of BTZ 
         black hole is also justified with the  exact partition function approach  \cite{zx}. 
     In fact, the corrections to entropy of black hole turned out to be  logarithmic   on  
Rademacher expansion of partition function   as well \cite{cv}.   In the case of
 string-black hole correspondence  \cite{ l,m,n,o},  the leading-order corrections to entropy 
 of black hole are  logarithmic terms. 
          Das et al. in Ref. \cite{hk} have
    shown that no  matter what kind of thermodynamic system  is dealt with, it always  leads 
    to the logarithmic correction.

     In recent past, the effects of small thermal fluctuation on the thermodynamics of small 
     black holes have been studied in great details.
      For instance, the  effect of quantum fluctuation on the thermodynamics of G\"odel black hole  was  studied in Ref. \cite{wer}.
         The leading-order corrections to the thermodynamics of Schwarzschild-Beltrami-de-sitter black hole \cite{hk1} and massive black hole  in
         Ads space \cite{hk2}  have been discussed.  In Ref. \cite{hk3},  
         the effect of statistical fluctuations on thermodynamics of dilatonic black hole has been found.  Again in \cite{nadeem} the 
         leading order order corrections to Horava Lifshitz black hole  were discussed and it was found that  the logarithmic corrections
         originate from  the thermal fluctuations and hence were interpreted as quantum loop corrections. The corrected thermodynamics of
         quasitopological black hole was examined in \cite{ul}. In particular the stability  and bound points for quasi topological
         black hole were studied and it was observed that  the correction terms with positive correction parameter resulted in more stable
          black hole while negative correction parameter introduces  instability in  such black holes. In \cite{islam} corrections to various
          thermodynamical quantities  of Vander walls black hole were evaluated. Also the effect of small statistical fluctuations on
          equation of state for Vander walls black hole has been discussed. Furthermore in addition to leading order corrections to entropy
          of small black holes (which are  found to be logarithmic in nature as expected), higher order  corrections  were also 
          analyzed   in  ref.\cite{sudhaker, sud4,sud5}. It  turned out that higher order corrections to entropy are inversely proportional to original
          entropy. In \cite{mir}  corrections to Reissner-Nordstrom, kerr, and charged Ads black holes were computed and in all these cases
          it was revealed that these corrections produce remnants. Moreover the effect of thermal fluctuations on various thermodynamical
          quantities  of a modified Hayward black hole was analyzed in \cite{faizal}. Also the effect of these correction terms on the first  law of thermodynamics  was examined.
          Although there are some discussion on the effect of thermal correction on the entropy and curvature of   BTZ black hole  \cite{qu}, the full discussion of such corrections on the thermodynamics of stationary BTZ black holes are not studied yet. This provides us an opportunity to bridge this gap. 
         
  Here we study the effect of thermal fluctuations  on the  thermodynamics of 
  non-rotating  BTZ black holes. In order to do so,  we first derive the leading-order logarithmic corrections due to thermal fluctuations to the entropy of stationary BTZ  black hole. In order to study the effect of this correction on  the entropy,  we do a comparative analysis by plotting   a graph between  corrected and non-corrected entropy with respect   to event horizon radius. From the plot, we find that the entropy of the system is positive valued. Also, there exists a critical horizon radius inside which the corrected entropy is decreasing function.
  We notice  that the correction parameter  does not
  play  a significant role for entropy of large sized black hole. However, for the entropy of small sized non-rotating 
  BTZ black hole, role of  the correction parameter turns out to be of paramount importance. 
       Furthermore, we derive the
  the first-order   correction  to the free energy. To analyze the modification due  to the thermal fluctuation, 
 we plot the corrected and non-corrected free energies with respect to the event horizon radius. We observe here that the Helmholtz free energy is decreasing function of radius. There exists a critical
horizon where thermal fluctuations do not affect the free energy. For larger black holes (with horizon
radius larger than critical horizon) the corrected free energy becomes less negative than its equilibrium
values. However, for smaller black holes, the correction terms make the free energy more negative.
 We also compute the leading-order corrections to the total mass (energy) of non-rotating BTZ black hole with the help of first law of thermodynamics and 
 plot a graph  for comparative analysis. 
  From the plot,  we see that total energy is monotonically increasing function of event horizon radius. The correction
term due to the thermal fluctuations decreases the total energy of the system.
We also derive the corrected pressure of BTZ black hole. 
Here we notice that for large sized black hole  the corrected
pressure coincides with the equilibrium pressure and becomes saturated. This means that for large black
holes the small thermal fluctuations do not affect the pressure considerably. However, for very small
black holes   the pressure increases to asymptotic  value due to the thermal
fluctuations. 
    We further derive the enthalpy of the system which is an increasing function of horizon radius for BTZ black holes. The correction terms due to thermal fluctuations decrease the
value of enthalpy. As long as the value of correction parameter increases the value of enthalpy decreases. We  also study the effects of thermal fluctuations on 
 Gibbs free energy which has zero value for the equilibrium case.  
   From the plot, we notice that there exist two critical points between which Gibbs free
energy is negative valued. Beyond the critical points,  Gibbs free energy is  positive valued. The corrected
Gibbs free energy with larger correction parameter   becomes more negative valued for smaller BTZ
black holes.  
 Finally,  we   discuss  the corrected specific heat due to thermal fluctuation. From the plot, we infer that specific heat is an increasing function 
    of horizon radius. The equilibrium value of
specific heat is positive always which suggests that the BTZ black holes are in stable phase. However, the thermal fluctuations make specific heat negative for
small black holes. This suggests that the black holes are in unstable phase due to thermal fluctuations
and the thermal fluctuations do not affect the stability of large sized black holes. we also calculate the  isothermal compressibility of the system under thermal fluctuations. We find that  in equilibrium state,  compressibility diverges which means that
system is highly compressible.   
    
    The paper is organized in the following way. In section II, we recapitulate the
    general expression for entropy under thermal fluctuation and compute it for the case of 
    non-rotating BTZ black holes. In section III, we derive various other equations of state 
    to describe the thermodynamics of black holes under the effect of thermal fluctuations.
    In section IV, we check stability of BTZ black holes under thermal fluctuations.
    Finally, we discuss results in the last section.
  \section{Logarithmic Corrections to entropy of non-rotating BTZ black hole}
Let us start discussion by writing metric for  the  non-rotating  BTZ black hole  with cosmological constant   $\Lambda= -\frac{1}{l^2}$ given as \cite{ban}
\begin{equation}
 ds^2=\left(\frac{r^2}{l^2}-8G_3M\right)dt^2+ \frac{dr^2}{\left(\frac{r^2}{l^2}-8G_3M\right)} +r^2d\theta^2,
\end{equation}
where $M$ refers to mass of black hole and  $G_3$ refers to the three dimensional Newton's gravitational constant. This metric  is of the form
\begin{equation}
  ds^2 = f(r)dt^2 +\frac{dr^2}{f(r)} + r^2d\theta^2,
\end{equation}
where metric function is given by
\begin{equation}
 f(r)=\frac{r^2}{l^2} - 8G_3M.
\end{equation}
Now, we compute Bekenstein-Hawking temperature from the above metric easily as following:
\begin{equation}\label{tem}
 T_H =\frac{f^{'}(r)}{4\pi}\arrowvert_{r=r_+} =\frac{r_+}{2\pi{l^2}}.
\end{equation}
The value of event-horizon radius $r_+$ can be obtained by setting
$
 f(r)=0$, 
which yields
\begin{equation}
 r_+ =\sqrt{8G_3M}{l}.
\end{equation}
The corresponding entropy for BTZ non-rotating black hole is given by \cite{ban}
\begin{equation}
 S_0 =\frac{\pi{r_+}}{2G_3}.\label{en}
\end{equation} 
In order to study the effect of thermal perturbations on the entropy of BTZ black hole, we first derive the exact expression  for the  entropy  of BTZ   black hole. 
In this regard, we write partition function describing  BTZ black hole as follows
      \begin{equation} 
       Z(\beta) =  \displaystyle\int_{0}^{\infty}{{dE}\rho{(E)}e^{-\beta{E}}},
      \end{equation}
where      $\beta =  \frac{1}{T_H} $         as Boltzmann constant $k=1$.
            Now, with the help of the inverse Laplace transform, one can  get the  density of states
      \begin{equation}
       \rho{(E)} =\frac{1}{2\pi{i}}{  \int_{{\beta}_0 - i\infty}^{{\beta}_0+ i\infty}{{d\beta}Z(\beta)e^{\beta{E}}}}   
        = \frac{1}{2\pi{i}}{  \int_{{\beta}_0 - i\infty}^{{\beta}_0+ i\infty}{d\beta }e^{S(\beta)}}.     \label{l}
        \end{equation}
      Here, 
  $ S(\beta) = \ln{Z(\beta)} + \beta{E} $  represents the exact entropy for the  black hole  and this depends on
      temperature explicitly. If one reduces the size of black hole  and expand  $S(\beta)$  around equilibrium, then using  the method
      of steepest descent (where $\frac{dS}{d\beta} = 0$ and $\frac{d^2S}{d{\beta}^2}  >  {0}$), we get
      \begin{equation}\label{eqp}
       S(\beta) = S_0 + \frac{1}{2}({\beta} - {\beta}_0)^2\frac{d^2S}{d{\beta}^2}|
       _{{\beta} = {\beta}_0}  + \mbox{(higher   order  terms)},
      \end{equation} 
      where $S_0$ represents the equilibrium  (original) value  of entropy.
    By inserting (\ref{eqp}) in (\ref{l}),   we have
      \begin{equation}
       \rho{(E)} = \frac{e^{S_0}}{2\pi{i}}{\int{{d\beta}e^{\frac{1}{2}{({\beta} -
       {\beta}_0})^2{\frac{d^2S}{d{\beta}^2}}}}}.
      \end{equation}
       Upon solving this integral, we get 
      \begin{equation}
       \rho{(E)} = \frac{e^{S_0}}{\sqrt{2\pi{\frac{d^2S}{d{\beta}^2}}}}.
      \end{equation} 
       Eventually, this  leads to
          \begin{equation}
      S = S_0 -\ln({S_0{T_H}^2})^{\frac{1}{2}},
      \end{equation} or
  \begin{equation}
    S = S_0 -\frac{1}{2}\ln{S_0{T_H}^2}.
   \end{equation}

       Here, without loss of generality, we may replace the $\frac{1}{2}$ factor of second term    by a more
       general correction parameter $\alpha$. This is because  the coefficient of log term 
modifies when Hawking temperature has power-law dependence on the entropy of system.        
        Thus, the corrected entropy by incorporating small
fluctuations around thermal equilibrium is given by
\begin{equation}\label{gen}
 S = S_0 - \alpha{\ln(S_0{T_H}^2)}.
\end{equation}     
    Moreover, it should be noted that second term in 
the above equation (which is obviously logarithmic in nature) appears due to small statistical fluctuations   around the thermal equilibrium. 
In other words,  we can say that the second term represents the 
 leading-order corrections to entropy of   BTZ black hole.
 
Now, by inserting the values of  hawking temperature  (\ref{tem}) and Bekenestein entropy (\ref{en}) to the expression (\ref{gen}), we get the perturbed expression for entropy of
 non-rotating BTZ black hole as
\begin{eqnarray}\label{entr}
 S &=& \frac{\pi{r_+}}{2G_3} - \alpha{\ln{\frac{{r_+}^3}{8\pi{G_3}l^2}}},\nonumber\\
 &=& \frac{\pi{r_+}}{2G_3} - 3\alpha{\ln{r_+}} + \alpha{\ln{G_38\pi{l^2}}}.
\end{eqnarray}
Here, we observe  that the last two terms of above expression, which  are  the logarithmic in nature, represent   corrections to the entropy of stationary BTZ black hole. These corrections  appear due to the thermal fluctuations around   equilibrium. By setting  $\alpha =0$, one can  retain the equilibrium  entropy of the system. Here, we can make a comparative analysis between the corrected and uncorrected entropy   by plotting a graph between them with respect to the event horizon radius with different possible values of correction parameter.  
\begin{figure}[htb]
 \begin{center}$
 \begin{array}{c }
\includegraphics[width=80 mm]{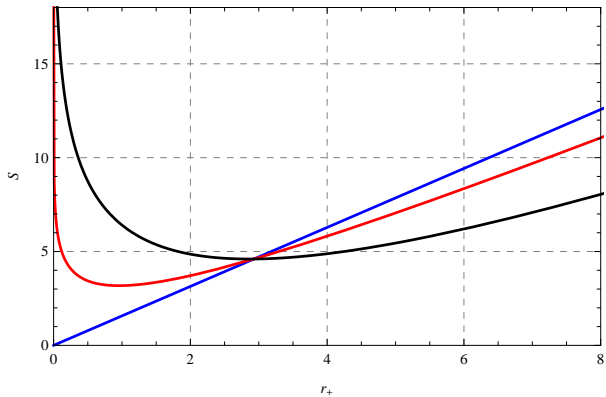}  
 \end{array}$
 \end{center}
\caption{Entropy  vs. the black hole horizon for $G_3$ and $l=1$.   Here, $\alpha=0$ denoted by blue line,
$\alpha=\frac{1}{2}$ denoted by red  line,   and $\alpha=\frac{3}{2}$ denoted by black line.}
 \label{fig1}
\end{figure}
From the FIG. \ref{fig1}, it is evident that the entropy of the system is positive valued.
 Also, there exists a critical horizon radius  below which    the  corrected entropy is decreasing function. However, similar to the equilibrium entropy behavior, the corrected entropy 
is an increasing function for the BTZ black holes with larger horizon radius than the critical value. This hints an important point that thermodynamics
of large sized black holes is not much affected by the  small thermal fluctuations as expected.  
\section{First-order corrected thermodynamic equation of states}
In this section, we would like to compute thermodynamical equation of states 
by incorporating small thermal fluctuation to the system. 
Once we have expressions of entropy and temperature, it is easy to compute various other
equations of states. For example, 
Helmholtz free energy (denoted by $F$) can be evaluated with the help of following formula:
\begin{equation}
 F = -\int{SdT_H}.\label{hem}
\end{equation}
By exploiting relations  (\ref{tem}), (\ref{entr})  and (\ref{hem}), we calculate the exact 
Helmholtz free energy for the BTZ black hole as follows,
\begin{equation}\label{free}
F= -\frac{r_+(12\alpha{G_3} + \pi{r_+} + \alpha{G_3}\ln{(8{G_3}l^2\pi)} - 12 \alpha{G_3}\ln{r_+})}{8G_3l^2\pi}
 \end{equation}
 \begin{figure}[htb]
 \begin{center}$
 \begin{array}{c }
\includegraphics[width=80 mm]{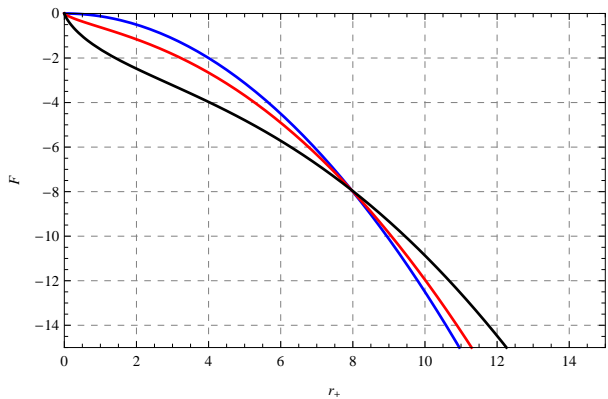}  
 \end{array}$
 \end{center}
\caption{Free energy   vs. the black hole horizon for $G_3=1$ and $l=1$.   Here, $\alpha=0$ denoted by blue line, 
$\alpha=\frac{1}{2}$ denoted by red  line,   and $\alpha=\frac{3}{2}$ denoted by black line.}
 \label{fig2}
 \end{figure}
 The behavior of Helmholtz free energy with respect to horizon radius can be seen from   FIG. {\ref{fig2}}. Here we see that the Helmholtz free energy shows a decreasing behavior 
 with increasing radius. There exists a critical horizon where thermal fluctuations do not affect the free energy. For larger black holes (with horizon radius larger than critical horizon)
  the corrected free energy becomes less 
 negative than its equilibrium values. However, for smaller black holes, the correction terms make  the free energy more negative.

The first law of thermodynamics  for uncharged stationary BTZ black holes  reads
 \begin{equation}
 {dE} ={T_HdS},
 \end{equation}
  which upon integration yields the energy
\begin{equation}
 E = \int{{T_H}d{S}}.\label{f},
\end{equation} where $E$ represents internal energy of the system.
Using the  values for $S$ (\ref{tem}), $T_H$  (\ref{entr}) and (\ref{f}), we get total energy by incorporating thermal fluctuations for
BTZ black hole 
 \begin{equation}\label{e}
 E = \frac{{r_+}^2}{8G_3l^2} - \frac{3\alpha{r_+}}{2\pi{l^2}}.
 \end{equation}
  \begin{figure}[htb]
 \begin{center}$
 \begin{array}{c }
\includegraphics[width=80 mm]{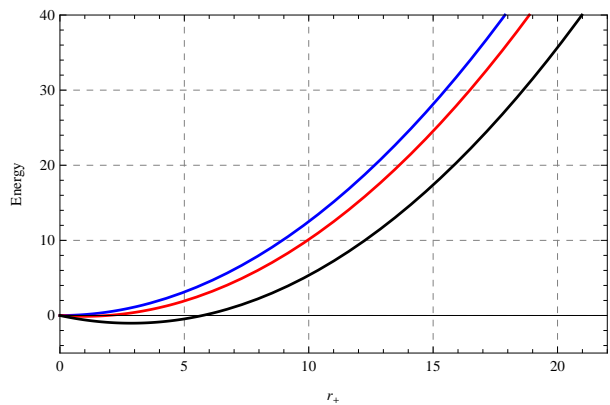}  
 \end{array}$
 \end{center}
\caption{Energy   vs. the black hole horizon for $G_3=1$ and $l=1$.   Here, $\alpha=0$ denoted by blue line, $\alpha=\frac{1}{2}$ 
denoted by red  line,   and $\alpha=\frac{3}{2}$ denoted by black line.}
 \label{fig3}
  \end{figure}
In order to do comparative analysis of corrected and equilibrium total energy, we  plot  FIG. {\ref{fig3}}. Here we observe that total energy is monotonically  increasing function of event horizon radius. The correction term due to the thermal fluctuations decreases the total energy of the system.  
  
From  the area-entropy theorem, we know  that the entropy of black hole  is proportional to
the  area covered by the event horizon. So, we can find the volume ($V$) of BTZ black hole as follows,
\begin{equation}
V  ={4G_3}\int{S_0dr_+}  = {\pi{r_+}^2}.\label{vol}
\end{equation}  

Since BTZ  black  holes are considered as the thermodynamic system, so one must be able to calculate other   macroscopic parameters such as pressure ($P$), which has the following standard definition in thermodynamics:
\begin{equation} \label{s1}
P = -\frac{dF}{dV} =-\frac{dF}{dr_+}\frac{dr_+}{dV}.
\end{equation}
So we need to find    $\frac{dF}{dr_+}$  and   $\frac{dV}{dr_+}$.   
 By plugging the values of (\ref{free}) and (\ref{vol}) in  (\ref{s1}), we obtain
   \begin{eqnarray}\label{p}
 P=\frac{\pi{r_+} + 2\alpha{G_3}\ln{8G_3l^2 \pi} - 6\alpha{G_3}\ln{r_+}}{8G_3l^2{\pi}^2{r_+}}
   \end{eqnarray}
  This expression  refers to exact pressure of BTZ black hole under the effect of small statistical fluctuations. 
   \begin{figure}[htb]
 \begin{center}$
 \begin{array}{c }
\includegraphics[width=80 mm]{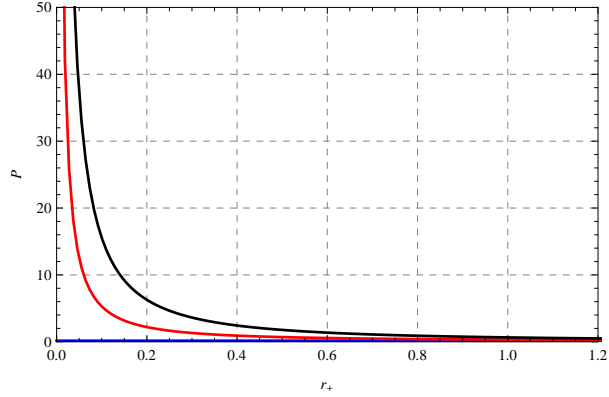}  
 \end{array}$
 \end{center}
\caption{Pressure  vs. the black hole horizon for $G_3=1$ and $l=1$.   Here, $\alpha=0$ denoted by blue line, $\alpha=\frac{1}{2}$
denoted by red  line,   and $\alpha=\frac{3}{2}$ denoted by black line.}
 \label{fig4}
\end{figure}
 We discuss the effect of the thermal fluctuations on the equilibrium pressure by  comparing 
the plot of   corrected and 
equilibrium    pressure with respect to event horizon radius.  
  
  From the  FIG. \ref{fig4}, it is obvious that for large sized black hole (i.e for large values of $r_+$)  the corrected  pressure coincides with the equilibrium pressure and becomes  saturated. This means that for large black holes the small thermal fluctuations do not affect 
  the pressure considerably. However, for very small black holes (i.e. for vanishingly small $r_+$) the pressure increases to the large value  due to the thermal fluctuations. So, remarkably,
 the contributions of  thermal fluctuations   to  the pressure of small BTZ black holes can not be ignored.  In fact, corrected  pressure  takes asymptotic values when horizon radius  tends to zero.

 The enthalpy ($H$) is an important   thermodynamic quantity, equivalent to the total heat content of a system, defined by
 \begin{equation}
 H = E +PV.
 \end{equation}
By using the  expressions of total energy (\ref{e}), pressure (\ref{p}) and  volume (\ref{vol}), the corrected enthalpy is calculated by
\begin{equation}
 H =\frac{r_+(-6\alpha{G_3} + \pi{r_+} + \alpha{G_3}\ln{8G_3l^2\pi} - 3\alpha{G_3}\ln{r_+})}{4G_3l^2\pi}.
\end{equation}
From this expression the behavior of enthalpy can be studied. 
  \begin{figure}[htb]
 \begin{center}$
 \begin{array}{c }
\includegraphics[width=80 mm]{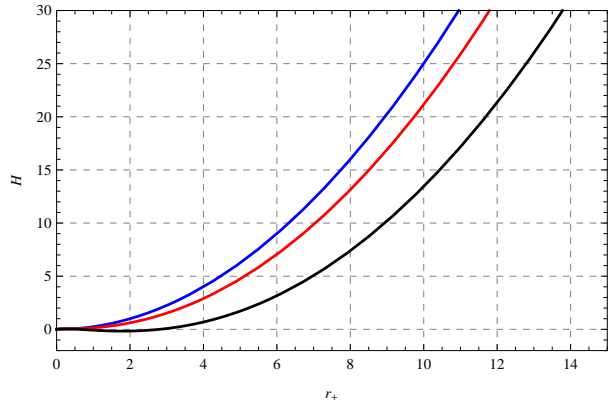}  
 \end{array}$
 \end{center}
\caption{Enthalpy  vs. the black hole horizon for $G_3=1$ and $l=1$.   Here, $\alpha=0$ denoted by blue line, $\alpha=\frac{1}{2}$
denoted by red  line,   and $\alpha=\frac{3}{2}$ denoted by black line.}
 \label{fig5}
 \end{figure}
 In order to study the effect of thermal fluctuations on the enthalpy, we plot a graph \ref{fig5}. From the plot, it is evident that the
 enthalpy  is an increasing function of horizon radius.   The correction terms due to thermal fluctuations decrease the  value of enthalpy. As long as the value of correction parameter 
 increases the value of enthalpy decreases. 

The Gibbs free energy  is the maximum amount of  work that can be extracted from a thermodynamically closed system. This maximum can be attained only in a completely reversible process. In order to compute 
  the corrected Gibbs free energy ($G$) for the BTZ black holes under the effect of 
  thermal fluctuations, we use following definition:
 \begin{equation}
 G = F +PV.
 \end{equation}
 Plugging the values of corrected Helmholtz free energy (\ref{free}), pressure   (\ref{p}), and volume (\ref{vol}) in the above expression,  we get the following expression for 
 first-order corrected Gibbs free energy:
 \begin{eqnarray}
 G = -\frac{\alpha{r_+}(6 +\ln{8G_3l^2\pi}  - 3\ln{r_+})}{4l^2\pi}.
 \end{eqnarray}
   \begin{figure}[htb]
 \begin{center}$
 \begin{array}{c }
\includegraphics[width=80 mm]{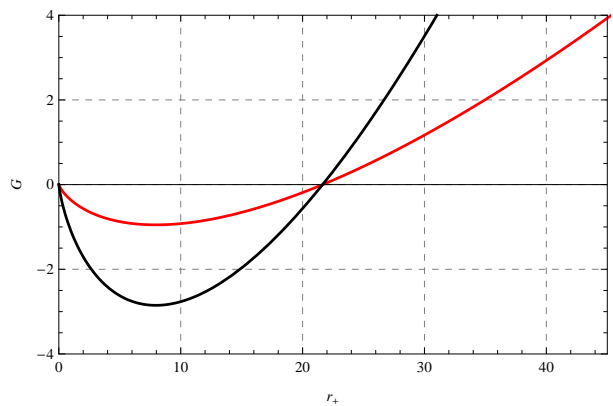}  
 \end{array}$
 \end{center}
\caption{Gibbs free energy  vs. the black hole horizon for $G_3=1$ and $l=1$.   Here,  $\alpha=\frac{1}{2}$ denoted by red  line, 
and $\alpha=\frac{3}{2}$ denoted by black line.}
 \label{fig6}
 \end{figure}
 In order to study the behavior of Gibbs free energy and the effects of thermal fluctuations, we plot a graph \ref{fig6}. 
From the figure,  we can see that the Gibbs free energy is zero for  BTZ black in thermal equilibrium. However,  when the 
 small statistical fluctuations  around equilibrium is considered, we get a non-zero 
 Gibbs free energy. We notice that there exist  two critical points between which Gibbs free energy  is negative valued. After the second critical value, it becomes positive valued. 
 The corrected Gibbs free energy with larger correction parameter $\alpha$ becomes more negative valued
 for smaller BTZ black holes (i.e. $r_+$ is less than the second critical point).  However, the corrected Gibbs free energy with larger correction parameter $\alpha$ becomes more positive valued
 for larger BTZ black holes.

\section{Stability}
 Now,  in order to discuss the stability of BTZ black hole, we just compute its specific heat
 by incorporating small thermal fluctuations.
 In  this regards, one can adopt two different
approaches. In first approach, the  specific heat  of the BTZ black hole will be calculated
and the positivity of the specific
heat will ensure the local thermal stability of the black holes.  
 In this case, the   unstable black holes
may go under phase transition.  In second  approach,  the thermal stability
can be investigated by calculating the determinant of Hessian matrix of mass with
respect to  extensive variables. Here, we will follow the first approach and will check the role of thermal fluctuations on the stability of black holes. 
By using the standard relation, the  specific heat ($C$) is calculated as    
  \begin{equation}
 C = \frac{dE}{dT_H} = \frac{dE}{dr_+}\frac{dr_+}{dT_H}
 = \frac{\pi{r_+}}{2G_3} - \frac{3\alpha}{2}.
  \end{equation}
  \begin{figure}[htb]
 \begin{center}$
 \begin{array}{c }
\includegraphics[width=80 mm]{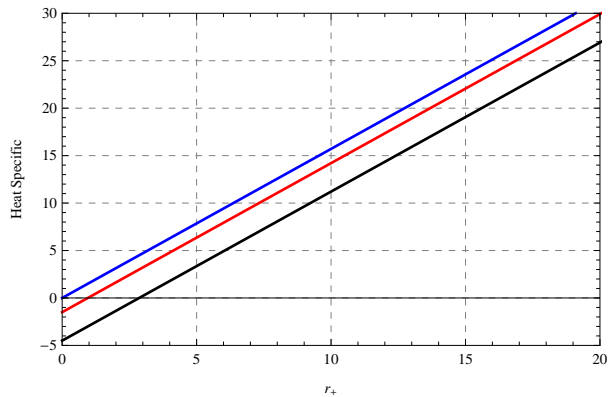}  
 \end{array}$
 \end{center}
\caption{Specific heat  vs. the black hole horizon for $G_3$ and $l=1$.   Here, $\alpha=0$ denoted by blue line, $\alpha=\frac{1}{2}$ 
denoted by red  line,   and $\alpha=\frac{3}{2}$ denoted by black line.}
 \label{fig7}
  \end{figure}
 We plot a graph \ref{fig7} to study the behavior of specific heat with respect to horizon radius. 
  From the graph,  one can infer that specific heat is an increasing function of horizon radius. The equilibrium value of specific heat is positive always which suggests that
  the BTZ black holes are in stable phase  without any thermal fluctuations.    However, the 
  thermal fluctuations cause   negative values to the specific heat for small black holes.
  This suggests that the black holes are in unstable phase due to thermal fluctuations and  
  the thermal fluctuations do not affect the stability of large sized black holes.  

 Now, let us check the effects of thermal fluctuations  on the isothermal compressibility ($K$) of uncharged  
  non-rotating BTZ 
  black hole.  The isothermal compressibility ($K$)   measures the respective change in volume due to    change in pressure. This is calculated by
 \begin{eqnarray}
   K = -\frac{1}{V}\frac{dV}{dP} =-\frac{1}{V}\frac{dV}{dr_+}\frac{dr_+}{dP}= \frac{8{\pi}^2{l^2r_+}}{\alpha(3 + {\ln{l^28\pi{G_3}}} - 3\ln{r_+})}.
 \end{eqnarray} 
 From the resulting expression, it is obvious  that when  $\alpha$ takes
 zero value the $K$ diverges which means that system  is highly compressible. Consequently,
 we get the zero value of  bulk modulus  as the compressibility is inverse to the
  bulk modulus.
 \section{Conclusion}
In this work,  we have investigated the effect of  statistical fluctuations on the small sized 
 BTZ black holes. We have found that the thermal fluctuations around equilibrium lead to a
  (logarithmic)  correction  to the entropy of stationary uncharged BTZ black holes. We have
  plotted a graph between entropy and horizon radius to discuss  the effect of thermal 
  fluctuations on the entropy of non-rotating BTZ black hole.   We have found that the entropy 
  takes   positive values always and there exists a critical point below which the corrected 
  entropy is decreasing function.
However,  the corrected entropy is an increasing function for
the large sized  BTZ black holes (for which horizon radius is larger than this critical 
value).

Moreover, we have calculated the first-order corrected Helmholtz free energy for the
 BTZ black hole where corrections appear due to the thermal fluctuations. We have checked the
 behavior of Helmholtz free energy and done comparative analysis between uncorrected and 
 corrected free energies analytically  by plotting a graph. We noticed that  
 the Helmholtz free energy is a decreasing function of horizon radius. 
 There exists
a critical point where the free energy is unaffected by  the thermal fluctuations. For large  black 
holes (with
horizon radius larger than  the critical horizon),  the corrected free energy becomes less negative  with respect
to its equilibrium values. However, for smaller black holes, the correction terms make  the free energy
more negative.
 
  We have evaluated the  corrected values of total energy (mass) to BTZ black hole with the help of first law of thermodynamics.We have found that total mass of the system  is a monotonically increasing function of event horizon radius. The correction
term due to the thermal fluctuations decreases the total mass of the system.
 We have calculated the volume and pressure of the system also. 
From the plot, we have observed  that for large sized black hole   the corrected
pressure coincides with the equilibrium pressure and becomes saturated. However, for very small
black holes (i.e. for vanishingly small horizon radius) the pressure takes asymptotically large values due to the thermal
fluctuations.  In fact,  for large black
holes the small thermal fluctuations do not affect the pressure considerably as expected. Remarkably, the contributions of thermal fluctuations to the pressure of small BTZ black
holes is considerable. In fact, the corrected pressure has become asymptotic large when horizon radius tends
to zero.
 
 Furthermore, we have calculated the
  corrected expression for the enthalpy. From the plot,  we have found that  the enthalpy is an
increasing function of horizon radius. The   thermal fluctuations decrease the
value of enthalpy.  The enthalpy decreases with the  larger values of correction parameter.
 The non-zero Gibbs free energy has   been found
  due to the thermal fluctuations. From the plot, we have noticed that  there exist two critical points. The    Gibbs free
energy is negative valued for small black holes. It takes positive values for large sized black holes. The corrected
Gibbs free energy with larger correction parameter has become more negative valued for smaller BTZ
black holes and become more positive valued for larger BTZ black holes. Here, we note that
the behavior of thermodynamics does not change corresponding to different values of 
correction parameter.

Finally,  we have derived the corrected expression of 
 specific heat for BTZ black hole and found that specific heat is increasing function of event horizon radius. From the graph, it is observed that  the equilibrium value of
specific heat is always positive  which suggests that the BTZ black holes are in stable phase without any
thermal fluctuations. However, due to the thermal fluctuations, the specific heat has become negative for
small black holes. This suggests that the small black holes are in unstable phase due to thermal fluctuations. In fact, the larger  correction parameter (taken arbitrarily) makes small black holes more unstable.   We have calculated the compressibility of the system, which diverges in equilibrium state which means that system  is highly compressible in equilibrium state. 
This leads to the zero bulk modulus  as the compressibility is inverse to the
  bulk modulus.

\end{document}